\documentclass[aps,prl,reprint,groupedaddress]{revtex4-1}

\usepackage{amsmath}
\usepackage{hyperref}
\usepackage{graphicx}
\usepackage{subfigure}
\usepackage{latexsym}
\usepackage[dvipsnames]{xcolor}
\usepackage{fullpage}
\usepackage{epstopdf}
\usepackage{dcolumn}
\usepackage{bm}
\usepackage{ulem}
\usepackage{units}

\begin{document}

\title{Signatures of long range dipolar interactions in artificial square ice}

\author{O. Brunn,$^{1,2,3}$ Y. Perrin,$^{1}$ B. Canals,$^{1}$ and N. Rougemaille$^{1}$} 
\affiliation{$^{1}$ Univ. Grenoble Alpes, CNRS, Grenoble INP, Institut NEEL, 38000 Grenoble, France \\
$^{2}$ Institute of Scientific Instruments of the Czech Academy of Sciences, Kr\'alovopolsk\'a 147, 612 64 Brno, Czech Republic \\
$^{3}$ Institute of Physical Engineering, Brno University of Technology, Technick\'a 2, 616 69 Brno, Czech Republic}

\date{\today}

\begin{abstract}
Analyzing the magnetic structure factor of a field demagnetized artificial square ice, qualitative deviations from what would predict the square ice model are observed.
Combining micromagnetic and Monte Carlo simulations, we demonstrate that these deviations signal the presence of interactions between nanomagnets that extend beyond nearest neighbors.
Including further neighbor, dipolar-like couplings in the square ice model, we find that the first seven or eight coupling strengths are needed to reproduce semi-quantitatively the main features of the magnetic structure factor measured experimentally.
An alternative, more realistic numerical scenario is also proposed in which the ice condition is slightly detuned.
In that case as well, the features evidenced in the experimental magnetic structure factor are only well-described when further neighbor couplings are taken into account.
Our results show that long range dipolar interactions are not totally washed out in a field demagnetized artificial square ice, and cannot be neglected as they impact the magnetic correlations within or at the vicinity of the ice manifold. 


\end{abstract}

\maketitle

\section{Introduction and Motivation}
\label{intro}

Two-dimensional arrays of interacting magnetic nanostructures are now well-established model systems to explore the physics of highly frustrated magnets \cite{Nisoli2013, Rougemaille2019, Skjaervo2020}.
Complementing what can be done with chemically synthesized compounds \cite{book2009, book2013}, artificially made spin lattices offer a lab-on-chip approach: almost any kind of geometry can be designed \cite{Skjaervo2020, Gilbert2014, Farhan2017, Marrows2018, Ladak2019}, magnetic interactions can be tuned \cite{Perrin2016, Stamps2019, Farhan2019a, Massouras2020}, structural defects can be engineered \cite{Drisko2017}, thermal fluctuations are adjustable in the desired temperature range \cite{Kapaklis2014, Chioar2014a, Schiffer2019}, the spin degree of freedom can be controlled \cite{Louis2018, Fischer2018, Leo2018}, etc.
Combine with the capability of imaging spin configurations directly in real space, at the scale of a nanomagnet, artificial spin systems can be viewed as experimental simulators of frustrated magnetism.

\begin{figure}[h]
\centering
\includegraphics[width=5cm]{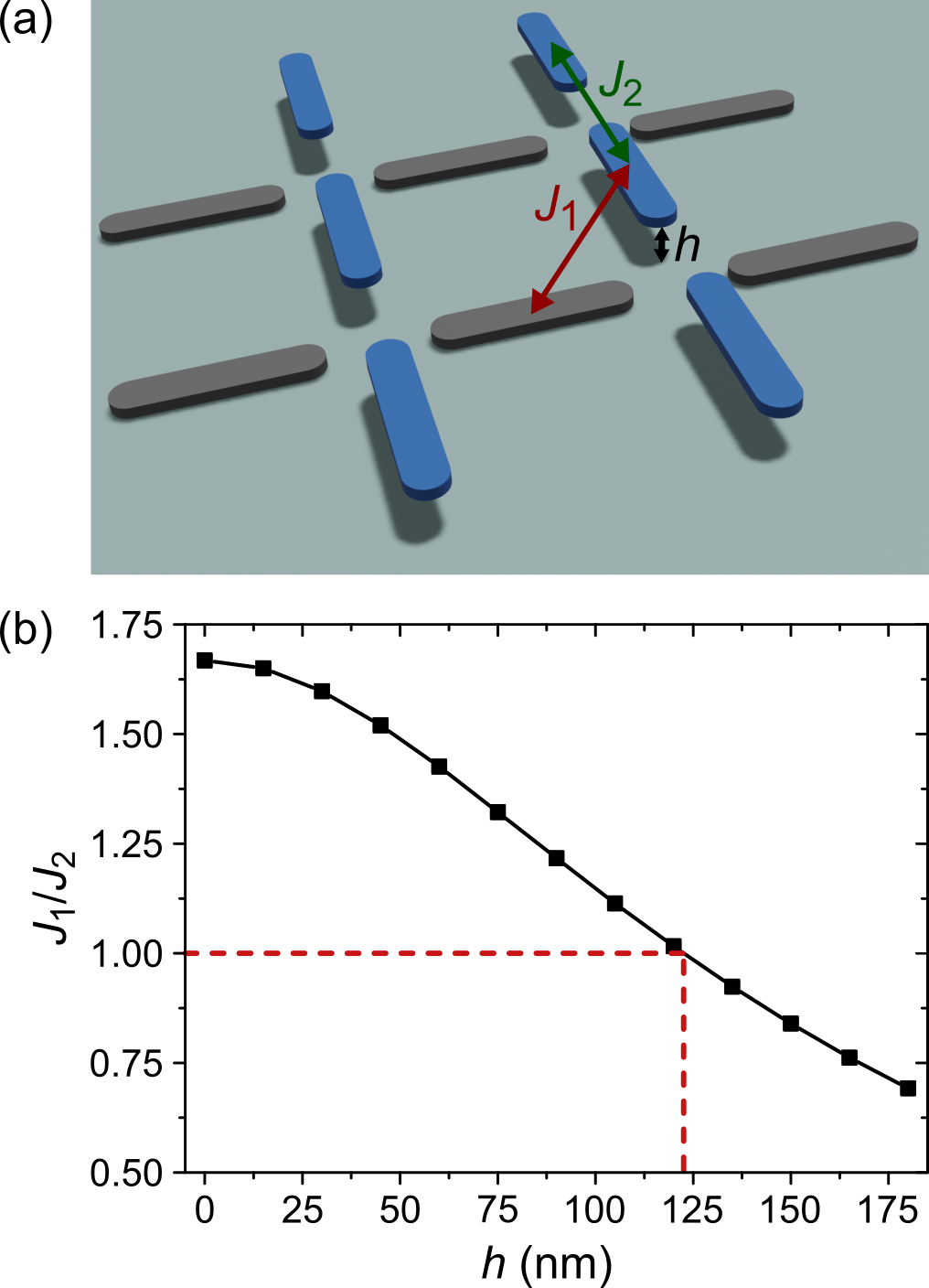}
\caption{a) Three-dimensional representation of a square lattice in which one of the two sublattices is shifted vertically by an offset $h$. The nearest-neighbor couplings $J_1$ and $J_2$ are indicated in red and green, respectively. b) Using micromagnetic simulations, the $J_1/J_2$ ratio is calculated as a function of the height offset $h$. For a critical (numerical) value $h_C^\textrm{num}$ close to 120 nm, $J_1/J_2=1$.
}
\label{fig1} 
\end{figure}

Among the works done so far, many studies on artificial spin systems have been devoted to the square geometry \cite{Wang2006, Morgan2011, Farhan2013, Porro2013, Kapaklis2014, Perrin2016, Ostman2018, Farhan2019b}. 
This geometry was first proposed to realize the so-called square ice model \cite{Lieb1967}, a model capturing the low-energy physics of water ice, but in two dimensions \cite{Nisoli2013, Rougemaille2019}.
However, two-dimensional square arrays of interacting nanomagnets do not show the expected disordered behavior: they order in an antiferromagnetic fashion because of the inequivalent coupling strengths between collinear and perpendicular nanomagnets \cite{Wang2006}.

Several modifications of the square lattice have been proposed, theoretically \cite{Moller2006, Chern2013, Perrin2019} and experimentally \cite{Gilbert2014, Perrin2016, Ostman2018, Farhan2019b}, to recover the ice degeneracy of the ground state.
This could be done, for example, by shifting vertically one of the two sublattices of the square lattice.
Doing so, the coupling strength $J_2$ [shown in green in Fig.~\ref{fig1}(a)] between collinear nanomagnets remains unchanged, whatever the amplitude of the vertical shift $h$.
However, the coupling strength $J_1$ [shown in red in Fig.~\ref{fig1}(a)] between orthogonal nanomagnets is varied continuously  [see Fig.~\ref{fig1}(b)], and can be made even negligible for large shifts.
The vertical shift $h$ is thus an experimental knob one can play with to adjust $J_1$ at will, especially to reach the $J_1=J_2$ condition required in the square ice model.
This was done for both athermal and thermally active arrays of nanomagnets, and the extensive degeneracy of the ice manifold was recovered \cite{Perrin2016, Farhan2019b}. 

Generally, artificial spin systems consist in the arrangement of nanomagnets coupled through magnetostatics. 
Thus, they are dipolar systems by nature, and considering nearest-neighbor interactions only is an approximation.
In particular, it is now established that long range interactions change the physics drastically of the artificial kagome ice \cite{Qi2008, Moller2009, Rougemaille2011, Zhang2013, Chioar2014a, Brooks2014, Montaigne2014, Canals2016, Drisko2017} and artificial kagome Ising antiferromagnet \cite{Chioar2014b, Chioar2016, Hamp2018}, compared to the same systems in which only nearest-neighbor interactions are taken into account \cite{Wills2002, Zhang2012}.

We can then wonder whether the presence of further neighbor couplings can also be detected in artificial square ice.
In other words, the question we want to address here is whether long range interactions can be observed experimentally or if they are washed out, for example because of intrinsic disorder or an inefficient energy minimization protocol.
This is an important question as dipolar interactions are known to lift the extensive degeneracy of the square ice manifold, and to ultimately order the system \cite{Moller2006}.


To address this issue, we fabricated a square ice system using a technique we developed previously \cite{Perrin2016}.
Then, we field demagnetized our lattices multiple times, imaged the resulting magnetic configurations using magnetic force microscopy, computed the magnetic structure averaged over the different experiments, and carefully analyzed intensity profiles in peculiar wavevector directions.
Comparing these intensity profiles with those deduced numerically from Monte Carlo simulations, we conclude that long range coupling strengths are indeed visible experimentally.
In other words, long range couplings are not washed out in our lattices, and impact the magnetic correlations within or at the vicinity of the ice manifold. 


\begin{figure}[h]
\centering
\includegraphics[width=7.5cm]{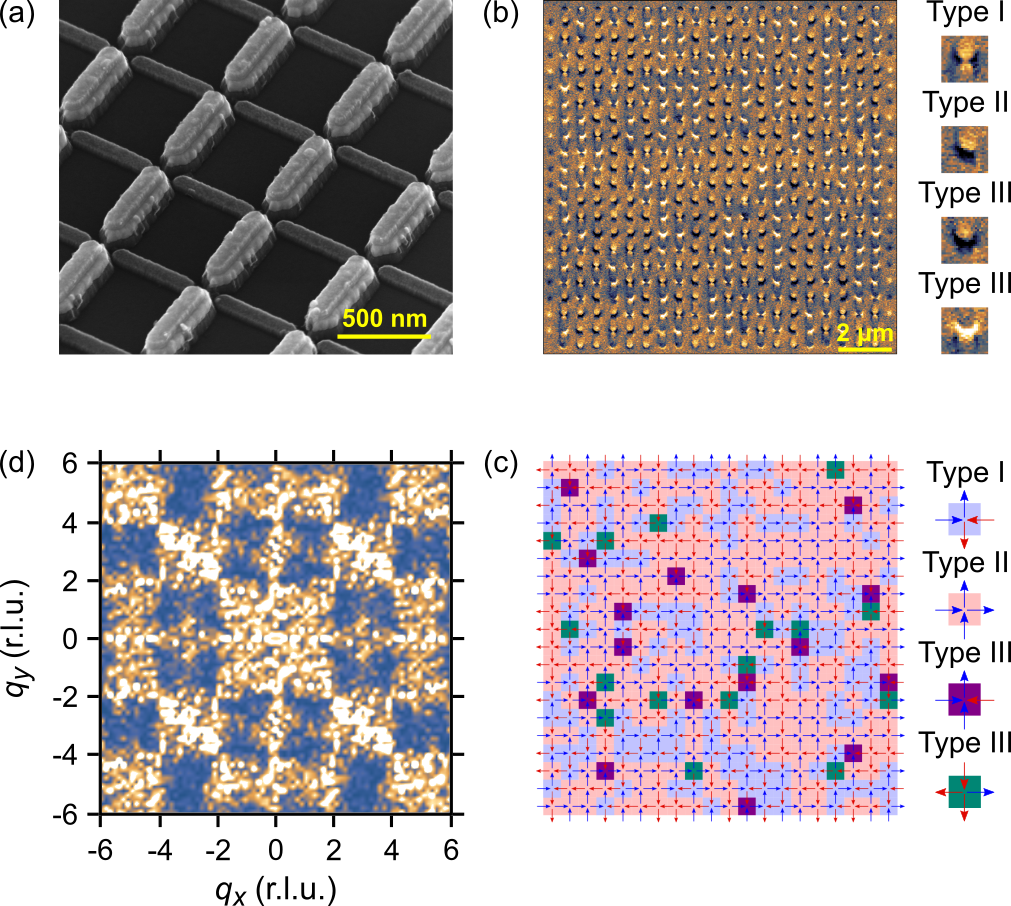}
\caption{a) Electron micrograph of an artificial square ice system in which one sublattice is shifted vertically. b) Magnetic image of a shifted square lattice. The black and white contrast allows unambiguous determination of the spin configuration. c) Spin configuration deduced from b) with a color code to highlight the different vertex types. d) Magnetic structure factor computed from c) This MSF ressembles the one of the square ice manifold.}
\label{fig2} 
\end{figure}

\section{Is the artificial square ice really short range?}
\label{asi}

The sample consists of square lattices made of 800 permalloy $500\times 100\times 30$ nm$^3$ nanomagnets, in which one of the two sublattices is shifted vertically [see Fig.~\ref{fig2}(a)].
The height offset is set to $h^\textrm{exp}=100$ nm, for which the liquid phase was already observed \cite{Perrin2016}.
The sample is demagnetized in a rotating, slowly decaying magnetic field.
After the demagnetization field protocol, the resulting magnetic configuration is imaged using magnetic force microscopy.
A typical image is reported in Figure~\ref{fig2}(b).
Analysis of the magnetic contrast at each vertex site allows the determination of the global spin configuration [see Fig.~\ref{fig2}(c)].
To characterize this spin configuration, the associated magnetic structure factor (MSF) is computed [see Fig.~\ref{fig2}(d)].

\begin{figure}[h]
\centering
\includegraphics[width=7.5cm]{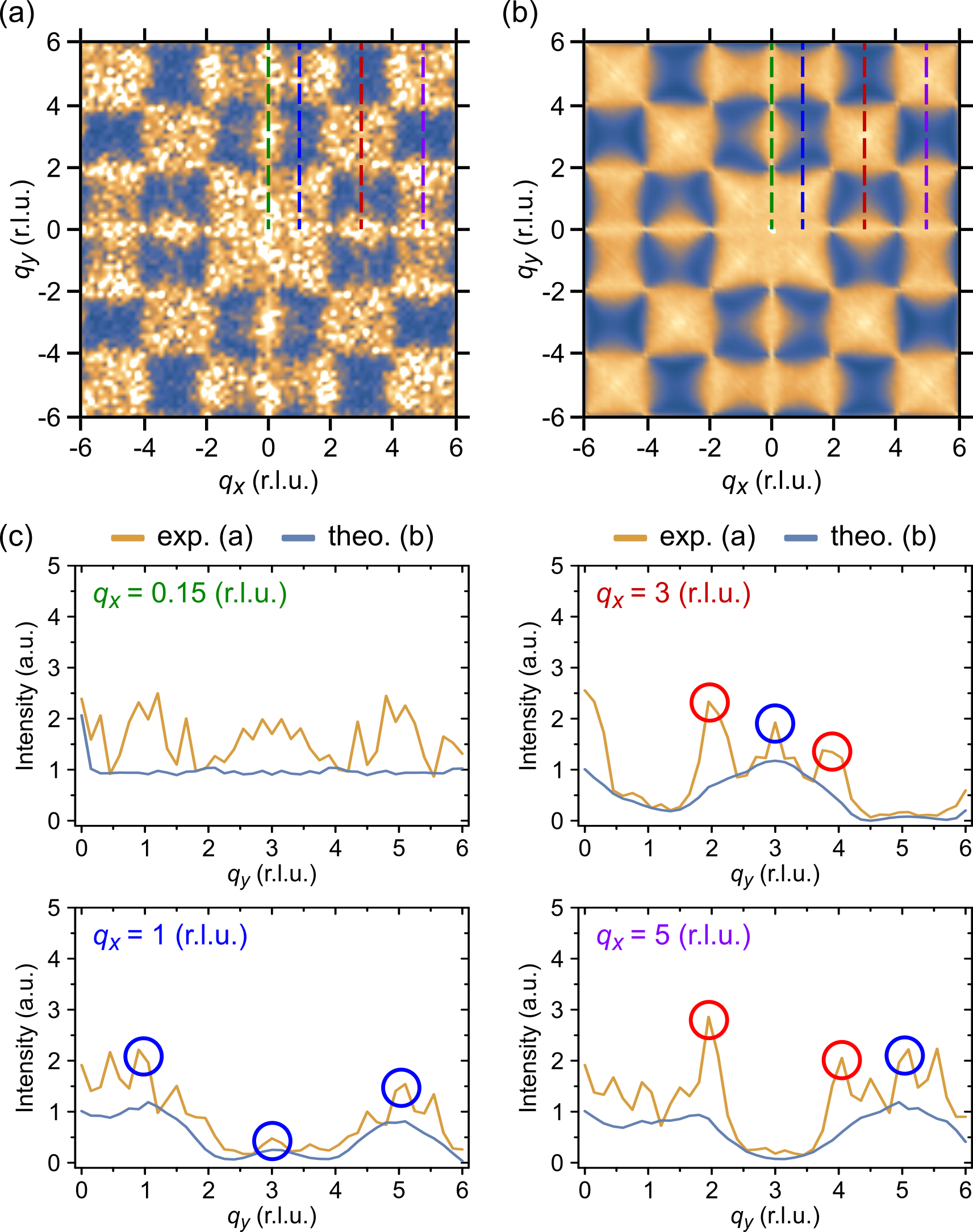}
\caption{(a) Experimental magnetic structure factor averaged over eight measurements (two lattices were demagnetized four times). (b) MSF corresponding to the low-energy manifold of the square ice model. The two MSF cover $\pm 6$ r.l.u. (reciprocal lattice units). (c) Intensity profiles obtained from (a) (in orange) and (b) (in blue) for the four $q$-directions $q_x = 0.15$, $1$, $3$ and $5$ r.l.u. The scan directions are shown in (a,b) as green ($q_x = 0.15$ r.l.u.), blue ($q_x = 1$ r.l.u.), red ($q_x = 3$ r.l.u.) and purple ($q_x = 5$ r.l.u.) dashed lines. The blue and red circles highlight specific wavevectors for which features are observed in the experimental MSF.}
\label{fig3}
\end{figure}

As reported previously \cite{Perrin2016}, the MSF shares all the features of the low-energy manifold of the square ice model: the background intensity is diffuse but structured, and exhibits emergent pinch points for certain wavevectors \cite{Perrin2016, Farhan2019b}.
Our measurements were reproduced several times, and a liquid-like state appears systematically in our lattices.
The MSF of eight magnetic images were then averaged to improve the statistics [see Fig.~\ref{fig3}(a)].
We note that the spin configurations always contain a significant fraction (about 8\%) of magnetic monopoles [see Fig.~\ref{fig2}(c)], i.e., local spin configurations having an energy higher than the one satisfying the so-called ice (or Bernal-Fowler) rule \cite{Bernal1933}.
The presence of magnetic monopoles reflects the fact that the arrested spin configurations are not in the ground state manifold.


At first sight, the physics we image in our lattices [see Fig.~\ref{fig3}(a)] strongly resembles the one of the celebrated square ice model [see Fig.~\ref{fig3}(b)].
However, careful inspection of the magnetic structure factor reveals features that cannot be accounted for by the square ice model.
For example, $q$ scans along $q_x \gtrsim 0$ in the average experimental MSF shows weak oscillations that are absent in the theoretical MSF [see Fig.~\ref{fig3}(c)].
One could think that this is a consequence of the poor statistics available experimentally or to a rough sampling in reciprocal space.
However, additional features also appear along the $q_x =$ 1, 3 and 5 directions, $q_x$ being expressed in reciprocal lattice units (r.l.u.).
For these directions, the average experimental MSF exhibits several peaks marked by red and blue circles in Figure~\ref{fig3}(c) (the meaning of the colored circles will be explained hereafter).
These peaks are absent in the theoretical MSF.
Our observations then show limitations when comparing the square ice model and its artificial realization.
The question we raise now is whether there is a simple, natural way to understand the origin of these features.
If so, the next question is whether we can identify a simple, realistic model that captures them all.
As we will see below, we may not provide a definitive answer, but we argue that these features originate from further neighbor interactions.

\section{Step 1: Estimate the coupling strengths}
\label{micromag}

\begin{table*}
\caption{Table providing the micromagnetic $J_i$($\mu$mag.) and point dipole $J_i$(dip.) coupling strengths for the first ten neighbors $i$. All values are normalized to $J_2$(dip.). The coupling strengths are calculated at $h^\textrm{exp} = 100$ nm, i.e., the experimental value of the critical height.}
\begin{ruledtabular}
\begin{tabular}{|c|c|c|c|c|c|c|c|c|c|c|}
$i=$ & 1 & 2 & 3 & 4 & 5 & 6 & 7 & 8 & 9 & 10 \\
\hline
$J_i$($\mu$mag.) & 2.220 & 1.933 & -0.333 & 0.045 & 0.104 & -0.052 & 0.132 & 0.073 & 0.058 & -0.018 \\
\hline
$J_i$(dip.) & 2.121 & 1.000 & -0.500 & 0.088 & 0.111 & -0.063 & 0.125 & 0.078 & 0.056 & -0.011 \\
\end{tabular}
\label{table1}
\end{ruledtabular}
\end{table*}

\begin{figure}[h]
\centering
\includegraphics[width=6cm]{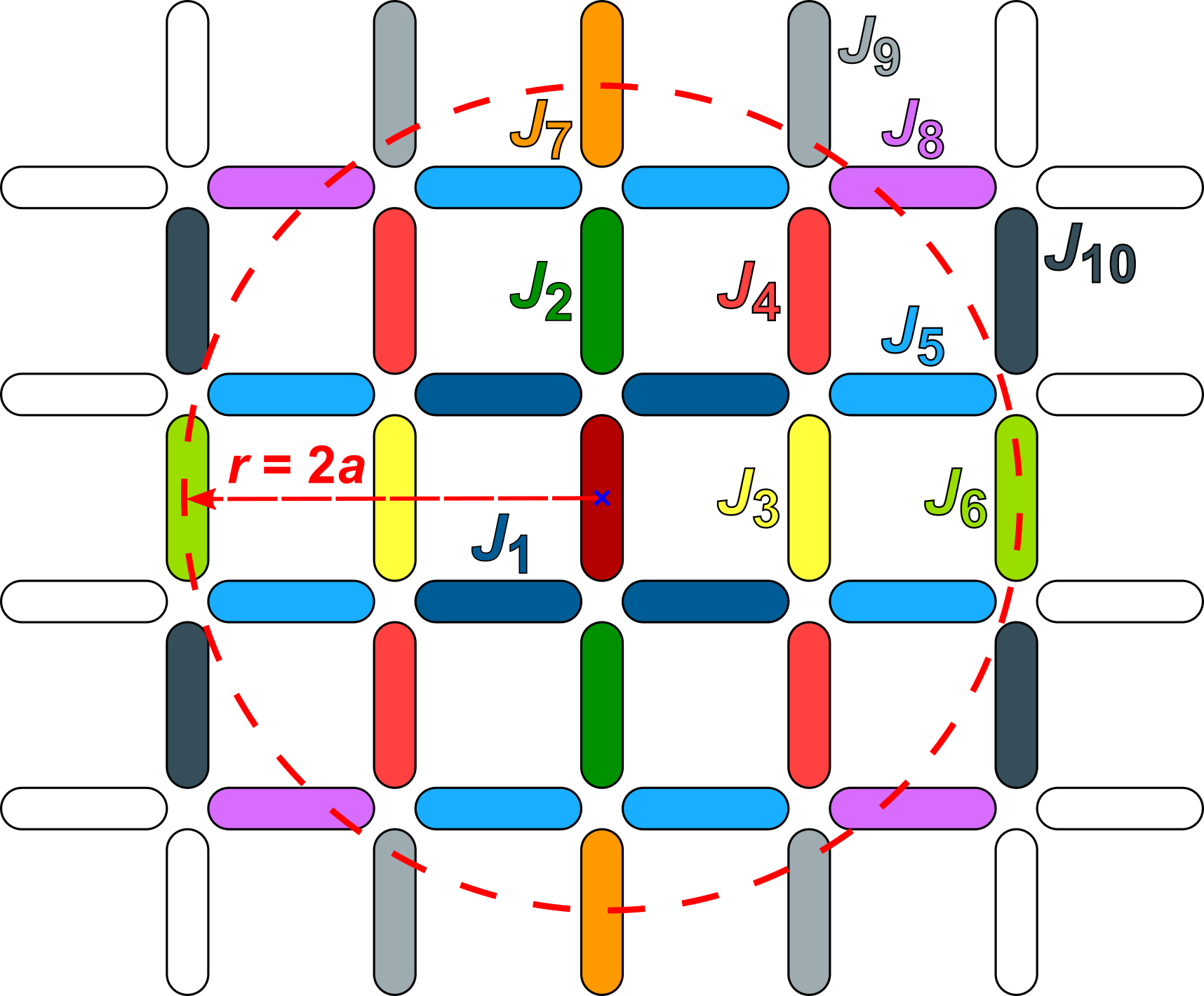}
\caption{Two-dimensional map showing the ten coupling strengths considered in this work. The circle in red has a radius of two lattice parameters $a$. Seven coupling strengths are included within this circle.}
\label{fig4}
\end{figure}

Although nanomagnets are often considered as Ising pseudospins, they are micromagnetic objects \cite{Rougemaille2013, Gliga2015, Dai2017, Paterson2019}.
Before computing the thermodynamic properties of the square ice model with further neighbor interactions, realistic values for the coupling strengths must be estimated.
To do so, we computed the micromagnetic energy of pairs of nanomagnets having the same dimensions as the ones fabricated experimentally ($500\times 100\times 30$ nm$^3$, with an edge-to-edge distance of 150 nm between collinear nanomagnets), and we considered the first ten neighbors (see Fig.~\ref{fig4}). 
In a vertically offset lattice, the $J_1$, $J_5$ and $J_8$ coupling strengths are derived from nanomagnets involving the two sublattices. 
For the other couplings, the two considered nanomagnets belong to the same sublattice.
The height offset $h$ then only matters when calculating $J_1$, $J_5$ and $J_8$.

The micromagnetic energies were computed using the OOMMF code from NIST \cite{OOMMF}. 
The mesh size was set to $1\times 1\times 15$ nm$^3$ to minimize finite difference effects.
We chose the material parameters commonly used for permalloy: the spontaneous magnetization $M_{s}$ is such that $\mu_{0}M_{s}$=1.0053 T, and the exchange stiffness is set to $A$=10 pJ/m.
Magnetocrystalline anisotropy is neglected.
The coupling strengths $J_i$ ($i = {1..10}$) are derived from the energy difference between a ferromagnetic and an antiferromagnetic configuration of the associated pair of nanomagnets.

First, we computed $J_1$ as a function of the vertical shift $h$ for values ranging from 0 to 180 nm, by steps of 15 nm.
As observed previously \cite{Perrin2016}, $J_1$ is larger than $J_2$ (which is used as a reference in the following and set to 1) when $h$ is smaller than a critical value $h_C^\textrm{num}$, and smaller otherwise.
This dependency of the $J_1$ coupling strength with the vertical shift $h$ is reported in Figure~\ref{fig1}(b).
A critical value $h_C^\textrm{num} \approx$ 120 nm is found, consistent with previous estimates \cite{Perrin2016}.

Regarding the other coupling strengths, the results from the micromagnetic simulations are reported in Table~\ref{table1}.
For comparison, the same coupling strengths are calculated within a point dipole approximation.
As expected, the coupling strength decreases quickly with the distance separation.
Except for the nearest neighbors, the values found using micromagnetic simulations are in fair agreement with those derived from a point dipole description.
Consistent with other works \cite{Rougemaille2011}, the $J_1$ and $J_2$ couplings are significantly larger because of the elongated shape of the nanomagnets.
Although some of the coupling strengths differ substantially in the two approaches (for example, $J_3$, $J_4$ and $J_{10}$), using those derived from micromagnetic simulations $J_i$($\mu$mag.) or those calculated from the point dipole approximation $J_i$(dip.) does not seem to affect the main results presented in the next sections.

\section{Step 2: Calculate the magnetic structure factor}
\label{thermo}

\begin{figure}[h]
\centering
\includegraphics[width=7.5cm]{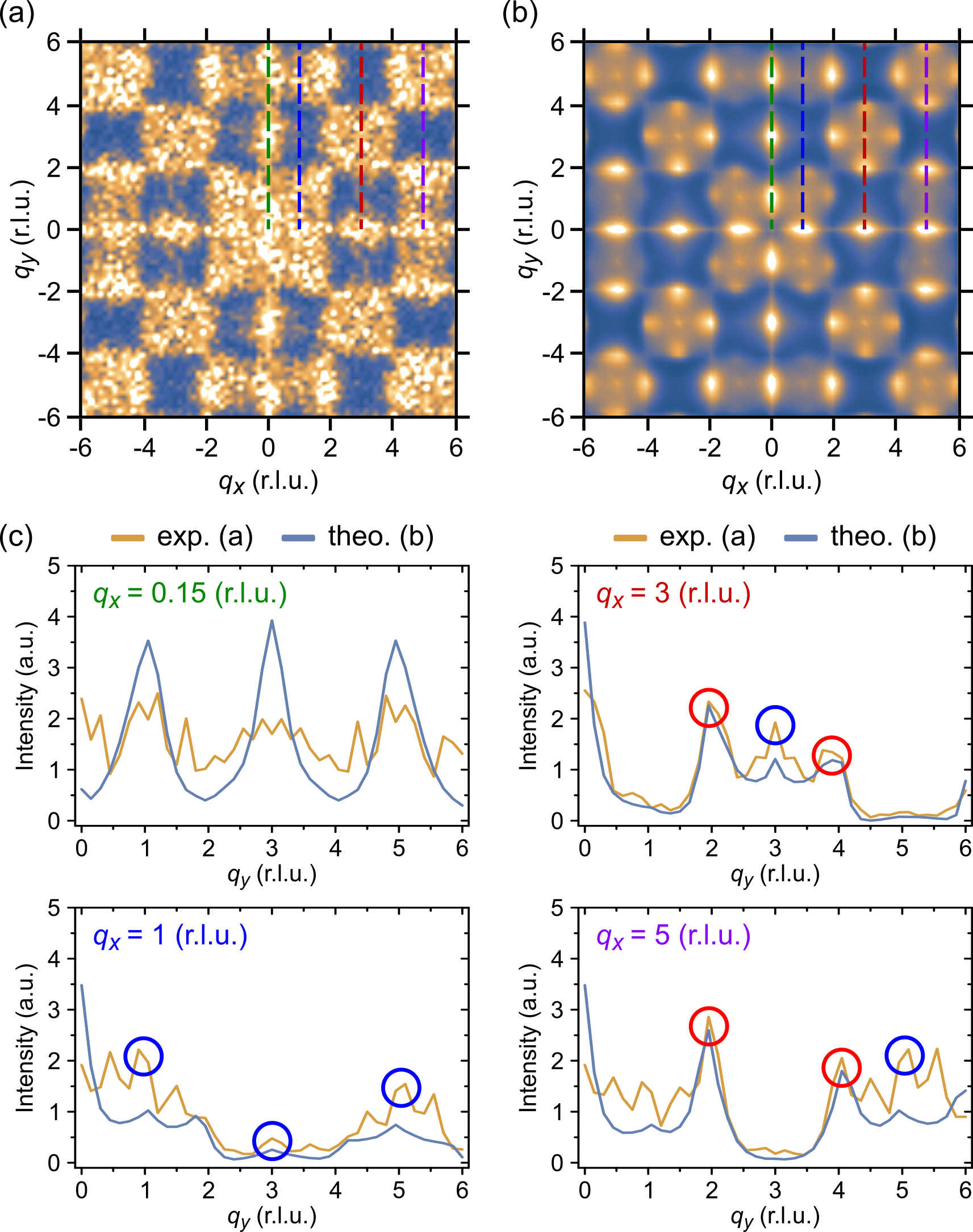}
\caption{(a) Experimental magnetic structure factor (MSF) averaged over eight measurements (two lattices were demagnetized four times). (b) MSF corresponding to the low-energy manifold of the spin Hamiltonian with the first seven $J$ couplings. The two MSF cover $\pm 6$ r.l.u. (reciprocal lattice units). (c) Intensity profiles obtained from (a) (in orange) and (b) (in blue) for the four $q$-directions $q_x = 0.15$, $1$, $3$ and $5$ r.l.u. The scan directions are shown in (a,b) as green ($q_x = 0.15$ r.l.u.), blue ($q_x = 1$ r.l.u.), red ($q_x = 3$ r.l.u.) and purple ($q_x = 5$ r.l.u.) dashed lines. The blue and red circles highlight specific wavevectors for which features are observed in the experimental MSF.}
\label{fig5}
\end{figure}

We now examine the thermodynamic properties of the spin model associated with the coupling strengths derived from the micromagnetic simulations.
To do so, we performed Monte Carlo simulations using the spin Hamiltonian:

\begin{equation}
H= - \sum_{i,j \mid r_{ij}<\alpha} J_{ij} \sigma_i . \sigma_j 
\label{Eq}
\end{equation}

\noindent where $\sigma_i$ and $\sigma_j$ are Ising variables on sites $i$ and $j$, separated by a distance $r_{ij}$, and $\alpha=5a\sqrt3/2$, $a$ being the lattice parameter.
The simulations were done for $12 \times 12 \times 2$ lattice sites \cite{note1} with periodic boundary conditions using a single spin flip algorithm. 
The cooling procedure starts from $T/J_1 = 100$ and ends when the dynamics freezes.
10$^4$ modified Monte Carlo steps (mmcs) are used for thermalization \cite{note2}.
Measurements follow the thermalization and are also computed with 10$^4$ mmcs.
The magnetic structure factor, composed of a matrix of $81 \times 81$ points covering an area of $\pm 6$ r.l.u. along the $q_x$ and $q_y$ directions in reciprocal space, is computed as a function of temperature.

To determine the origin of the features present in the experimental MSF (see Fig.~\ref{fig3}), we proceed as follow:
1) We assume that the square ice condition is obeyed: $J_1 = J_2 = 2$. 
2) The thermodynamic properties of the spin Hamiltonian are computed by incorporating the eight other coupling strengths one at a time. 
A first simulation is performed with $J_1$ and $J_2$ only, a second one is performed after adding $J_3$, a third one after adding $J_4$ to the three other values, etc, until all ten $J_i$ values reported in Table~\ref{table1} are considered.
3) For all these simulations, intensity profiles along the $q_x = 0.15, 1, 3$ and 5 r.l.u. directions are compared to the experimental ones, similar to what is reported in Fig.~\ref{fig3} for the square ice.
4) For each spin Hamiltonian, we determine the temperature that best fits the data.

\begin{figure}[h]
\centering
\includegraphics[width=7.5cm]{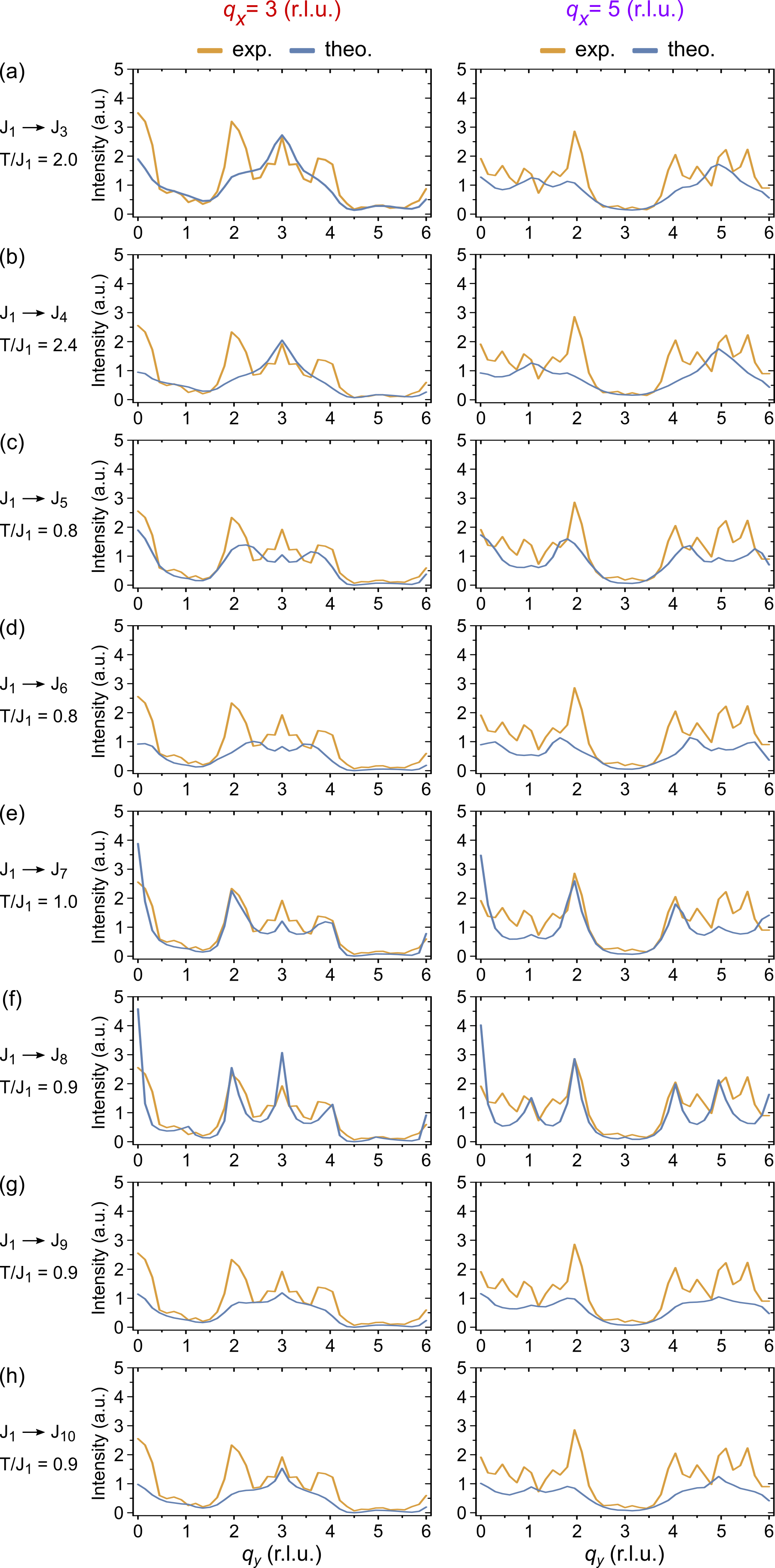}
\caption{Intensity profiles in the $q_x = 3$ and $5$ r.l.u. directions obtained from the magnetic structure factors computed for the spin Hamiltonians with interactions up to the third (a), fourth (b), fifth (c), sixth (d), seventh (e), eighth (f), ninth (g) and tenth (h) coupling strengths. The theoretical scans, shown in blue, are the best fits to the experimental data (orange scans). The temperature associated to these scans is provided.}
\label{fig6}
\end{figure}

Qualitative, and even sometimes semi-quantitative, agreement is found with the experiments when the Hamiltonian includes the first seven or first eight coupling strengths, whereas the other simulations fail to capture all the features observed in the average MSF (see Figs.~\ref{fig5} and \ref{fig6}).
We then conclude that the features observed in the intensity profiles [within the colored circles in Fig.~\ref{fig3}(c) and the oscillations at $q_x = 0.15$ r.l.u.] are not statistical noise or artefacts, but a real signal.
We also conclude that interactions up to 7 or 8 neighbors must be taken into account to reproduce our observations (see Fig.~\ref{fig5}).
Assuming the ice condition is obeyed in our artificial lattices, the typical range of the dipolar interaction we are able to probe experimentally is then about two lattice parameters (see dashed red circle in Fig.~\ref{fig4}).

\section{Discussion}
\label{discussion}

Based on our findings, we now briefly address the following questions:

\noindent 1) What is the origin of the peaks marked by red and blue circles in Figs.~\ref{fig3} and \ref{fig5}? How are they linked to the real space spin configurations?

\noindent 2) Is our model realistic, or is there an alternative, maybe more relevant scenario that also captures the features observed in the experimental MSF? In that case, does it need to include long range couplings as well?

\noindent 3) Should we expect our artificial system to ultimately order when properly demagnetized, or can we consider it as a good approximation of the square ice?

\subsection{The features of the MSF}

\begin{figure}[h]
\centering
\includegraphics[width=7.5cm]{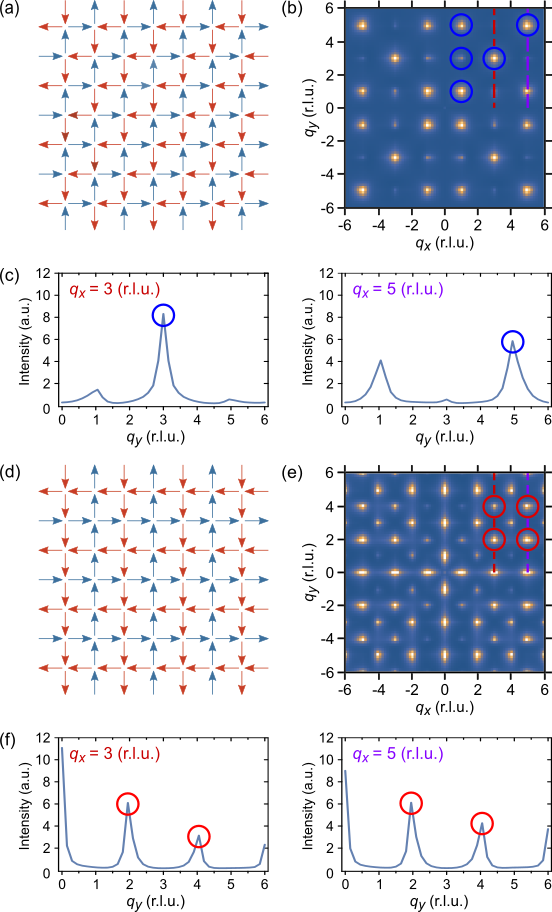}
\caption{(a) Real space and (b) reciprocal space representation of the ordered ground state of the Rys-F model. (c) Intensity profiles obtained from (b) for the $q_x = 3$ and $5$ r.l.u. directions. (d) Real space and (e) reciprocal space representation of the ordered ground state of the Slater-KDP model with an antiferromagnetic coupling between adjacent lines. (f) Intensity profiles obtained from (e) for the $q_x = 3$ and $5$ r.l.u. directions. The two  MSF shown here are obtained at finite temperature to make the Bragg peaks broader). The blue and red circles highlight specific wavevectors for which features are observed in the experimental MSF.}
\label{fig7}
\end{figure}

As mentioned above, we might wonder what is the origin of the features marked by colored circles in Figs.~\ref{fig3} and \ref{fig5}, i.e., what is the nature of the associated spin-spin correlations. 
To answer that question it is instructive to remember that dipolar interactions lift the ice degeneracy in a system with a height offset like ours \cite{Moller2006}.
However, Monte Carlo simulations show that the ground state depends on the value of this height offset \cite{Moller2006}.
When the height offset $h$ is smaller than a critical value $h_C$, the ground state is ordered and antiferromagnetic in the sense of the Rys-F model \cite{Rys1963, Lieb1967d}.
The associated ground state configuration and MSF are represented in Figs.~\ref{fig7}(a-c). 
When $h>h_C$, the ground state is also ordered and antiferromagnetic but in the sense of the Slater-KDP model \cite{Lieb1967b}: it then consists of an antiferromagnetic alignment of fully polarized lines [see Figs.~\ref{fig7}(d-f)].

Interestingly, we find that the magnetic Bragg peaks in these two ground states are located in reciprocal space precisely where the oscillations and spikes are measured in the experimental MSF [Fig.~\ref{fig5}(c)].
For example, the Rys-F ground state leads to Bragg peaks at $(q_x,q_y)=(3,3)$ and $(5,5)$.
This fits well with the features highlighted by blue circles in Figure~\ref{fig5}(c).
Similarly, the other possible ordered ground state leads to Bragg peaks at $(q_x,q_y)=(3,2)$, $(3,4)$, $(5,2)$ and $(5,4)$.
This also corresponds to the locations highlighted by red circles in Figure~\ref{fig5}(c).
We note that even the asymmetry between the two peak intensities marked by a red circle in Fig.~\ref{fig5}(c) at $q_x = 3$ and $q_x = 5$ r.l.u. is found in the MSF of the ground state.
In fact, the same argument holds as well for the Bragg peaks originating from the Rys-F model [the $(q_x,q_y)=(3,3)/(5,5)$ and $(q_x,q_y)=(1,1)/(1,5)$ peaks are more intense than the $(3,5)$ and $(1,3)$ ones, respectively, like we find in the experiments].
Finally, we note that the weak oscillations we observe in our artificial lattices near the $q_x = 0$ direction are also captured by the ground state of the Slater-KDP model [see Fig.~\ref{fig7}(e)].

The two antiferromagnetic ground states considered here allow the description of all the features evidenced in our experiments.
Their occurence in our lattices thus indicates the presence of extra spin-spin correlations on top of the square ice manifold.
In other words, the magnetic configurations resulting from the demagnetization of our arrays are not exactly a random arrangement of type I and type II vertices.
Instead, type I vertices are slightly more surrounded by other type I vertices than they should. 
Type II vertices belonging to adjacent lines in the square lattice are also slightly more antiferromagnetically coupled than expected in the square ice.

\subsection{An alternative scenario}

Although our measurements and the model developed above show semi-quantitative agreement when considering coupling strengths up to $J_7$ or $J_8$, this agreement might seem surprising. 
First, the model assumes that the square ice condition is fulfilled, i.e., $J_1=J_2$.
If one can reasonably consider that the ice condition is approached experimentally with $h=100$ nm, it is unlikely that is it strictly obeyed.
Moreover, a fair agreement is found only if the dipolar interaction has a cutoff radius of two lattice parameters.
If this radius is taken smaller or larger, agreement is lost (see Fig.~\ref{fig6}).
We then have to admit that either there is a kind of miracle that leads to an effective cutoff radius of the dipolar interaction of 2$a$ in our experiments, or that the agreement reported in Fig.~\ref{fig5} is fortuitous.  

We emphasize that considering $J_1$ and $J_2$ only is not sufficient to interpret our results.
We examined the case where $J_1 = J_2$ [see Fig.~\ref{fig3}], but the same conclusion is drawn if $J_1 \neq J_2$: all the features we observe cannot be described.
Assuming $J_1 > J_2$, the extra peaks within the blue circles in Figs.~\ref{fig3} and \ref{fig5} can be fitted, but the peaks within the red circles are missing.
Assuming $J_1 < J_2$, only the peaks within the red circles can be fitted.
Considering further neighbor couplings is thus mandatory.

\begin{figure}[h]
\centering
\includegraphics[width=7.5cm]{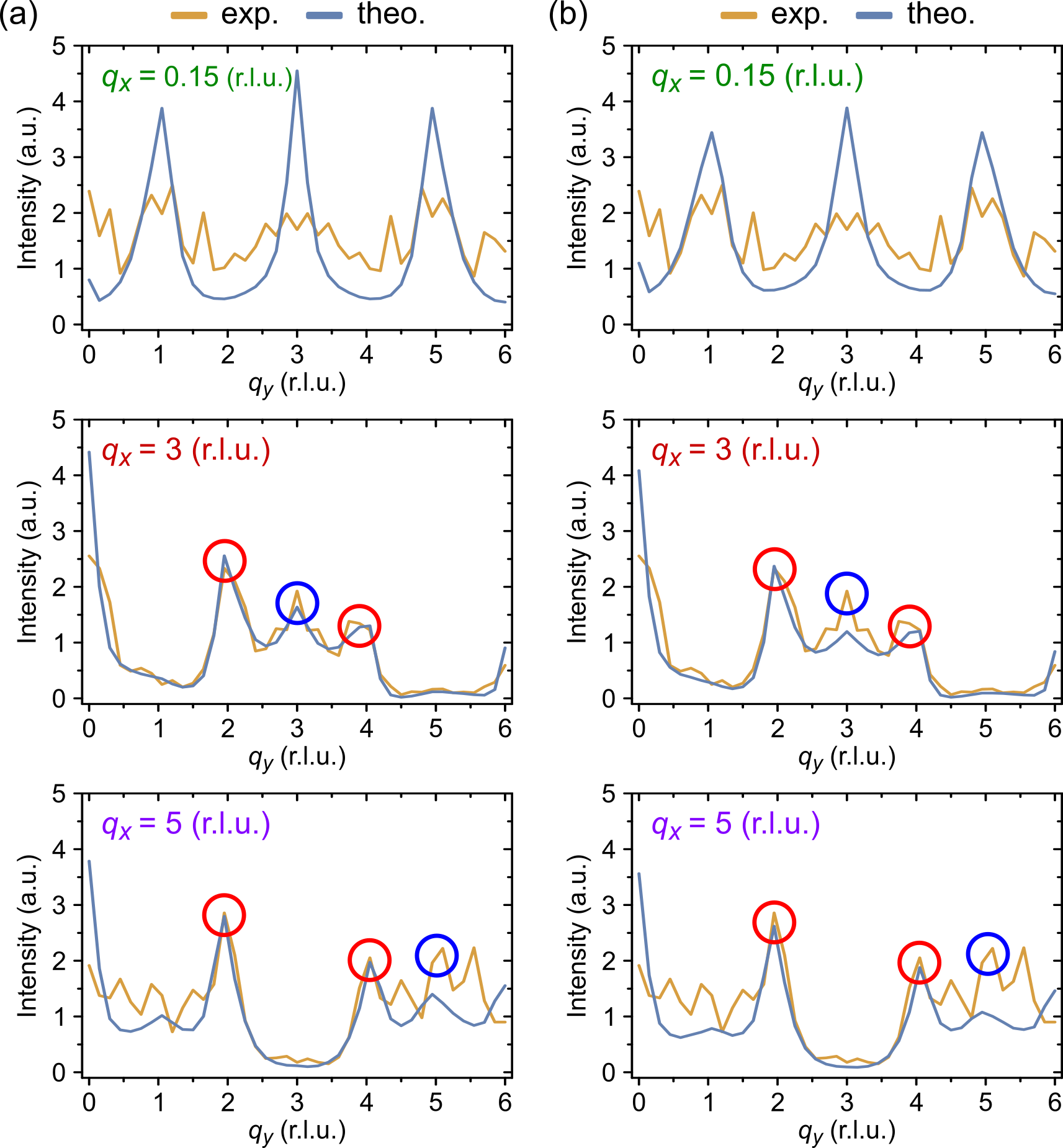}
\caption{Intensity profiles along the $q_x = 0.15, 3$ and $5$ r.l.u. directions for two spin models: (a) $J_1=2$, $J_2=2.1$, $J_3=-0.333$, and (b) $J_1=2$, $J_2=2.2$, $J_3=-0.333$, $J_4=0.045$. The temperature in the Monte Carlo simulations is $T=1.5 J_1$. The blue and red circles highlight specific wavevectors for which features are observed in the experimental MSF.}
\label{fig8}
\end{figure}

Including the coupling constants one at a time in our simulations is instructive to suggest an alternative, possibly more realistic scenario.
When the first three $J$ constants are incorporated into the spin Hamiltonian, the emergent Bragg peaks associated with the Rys-F model are favored, whereas those associated with the Slater-KDP model are absent [see Fig.~\ref{fig6}(a)].
An intuitive way to recover the missing peaks is to detune the $J_1 = J_2$ condition, while keeping the $J_3$ value (-0.33) unchanged.
In particular, as mentioned above, the $J_1 < J_2$ condition strengthen the correlations at $(q_x,q_y)=$(3,2), (3,4), (5,2) and (5,4).
Doing so, we indeed find again a semi-quantitative agreement when choosing $J_1=2$, $J_2=2.1$ and $J_3=-0.33$ [see Fig.~\ref{fig8}(a)].
In fact, the same approach works as well with $J_3$ and $J_4$ [see Fig.~\ref{fig8}(b)], and likely beyond. 

This scenario could be justified in our artificial lattices if the height offset $h$ was slightly higher than the critical value.
In that case, further neighbor couplings will induce the extra spin-spin correlations we measure experimentally.
In other words, even though the dipolar square ice model we first investigate is maybe not the most relevant model to consider, coupling strengths that extend beyond nearest neighbors must be taken into account to capture the physics we image.
Thus, we do observe signatures of long range dipolar interactions in our artificial square ice. 


\subsection{Ordering vs. dynamical freezing}

To conclude, we might wonder why our system does not order as the Monte Carlo simulations predict \cite{Moller2006}, and why the experimental MSF strongly resembles the one of the square ice, at least at first sight.
One reason, which is common to many frustrated magnets \cite{Melko2004}, is the freezing of the spin dynamics as the (effective) temperature of the system is reduced \cite{Schanilec2020}.
More specifically, the square ice becomes a loop model at low temperature.
Once the system entered the ice manifold, any single spin flip event breaks the divergence-free constraint, leading to the nucleation of a monopole - antimonopole pair.
Such an event requires an energy barrier to be overcome.
For sufficiently low temperatures, the nucleation of a monopole pair becomes statistically unlikely.
The only way to jump from one microstate to another is to reverse a chain of neighboring spins, i.e., to excite loop moves.
The single spin flip dynamics then freezes.
Whether artificial lattices are field demagnetized or thermally annealed, the spin dynamics is expected to freeze, even in an ideal, defect-free system.
Exploring the ground state manifold of an artificial ice magnet is thus a lost battle \cite{Rougemaille2019}, and reaching a microstate belonging to the ground state is challenging.
The arrested configurations usually obtained in artificial systems then remain at a relatively high effective temperature.
In that sense, it is not a surprise that our demagnetized arrays capture the physics of the square ice and do not order.




\bigskip

This work was supported by the Agence Nationale de la Recherche through project no. ANR-17-CE24-0007-03 'Bio-Ice', and its infrastructure partially supported by the CAS (RVO:68081731).





\begin{thebibliography}{}
\bibitem{note1}
We checked that our results do not depend on the system size. In specific cases, simulations were run for $12 \times 12 \times 2$, $20 \times 20 \times 2$ and $30 \times 30 \times 2$ lattice sites. No substantial difference was observed.
\bibitem{note2}
One modified Monte Carlo step corresponds to a sequence of spin flip attempts such that, on average, each spin is flipped once. In practice, this corresponds to a number of trials equals to N (number of sites) $\times$ the inverse of the acceptance rate (average value of the spin flip acceptance). At the lowest temperatures, $10^4$ mmcs corresponds typically to $10^8$ spin flip trials. This allows to achieve statistical independence.
\end{thebibliography}
\end{document}